\documentclass[12pt,titlepage]{article}
\usepackage{amssymb}
\usepackage{epsfig}
\usepackage{bm}
\font\grande=cmr10 scaled \magstep4
\font\medio=cmr10 scaled \magstep2
\outer\def\beginsection#1\par{\medbreak\bigskip
      \message{#1}\leftline{\bf#1}\nobreak\medskip
\vskip-\parskip
      \noindent}

\newcommand{\eq}{\begin{equation}}

\newcommand{\eqx}{\end{equation}}
\newcommand{\eqn}{\begin{eqnarray}}

\newcommand{\bi}{\begin{itemize}}

\newcommand{\eqnx}{\end{eqnarray}}
\newcommand{\ei}{\end{itemize}}
\newcommand{\ad}{{a^{\dagger}}}
\newcommand{\fd}{f^{\dagger}}
\newcommand{\bd}{b^{\dagger}}

\newcommand{\Qd}{{Q^{\dagger}}}

\newcommand{\nn}{\nonumber}
\newcommand{\ra}{\rangle}
\newcommand{\la}{\langle}

\begin{document}
\titlepage

\begin{flushright}
\vspace{5mm}
CERN-PH-TH/2006-185 \\
TPJU-10/2006
\end{flushright}
\vspace{10mm}
\begin{center}

\grande{A supersymmetric matrix model:  \\
\vspace{2mm} III.  Hidden SUSY in statistical systems}

\vspace{15mm}

\large{G. Veneziano}

\vspace{2mm}

 {\sl Theory Division, CERN, CH-1211 Geneva 23, Switzerland }

{\sl and}

{\sl Coll\`ege de France, 11 place M. Berthelot, 75005 Paris, France}
\vspace{10mm}

   \large{J. Wosiek}

   \vspace{2mm}

   {\sl M. Smoluchowski Institute of Physics, Jagellonian University}

{\sl Reymonta 4, 30-059 Cracow, Poland}

\vspace{8mm}

\end{center}

\centerline{\medio  Abstract}
\vskip 5mm
\noindent
The Hamiltonian of a recently proposed
 supersymmetric  matrix model has been shown to become block-diagonal  in the large-$N$, infinite 't Hooft coupling  limit. We show that (most of) these blocks can be mapped into seemingly non-supersymmetric  $(1+1)$-dimensional statistical  systems, thus implying non-trivial (and apparently yet-unknown) relations within their spectra.  Furthermore, the ground states of XXZ-chains with an odd number of sites and  asymmetry parameter $\Delta = - 1/2$,  objects of  the much-discussed Razumov--Stroganov conjectures, turn out to be just the strong-coupling supersymmetric vacua of  our matrix model.

  \vspace{5mm}

\begin{flushleft}
CERN-PH-TH/2006-185 \\
TPJU-10/2006\\
September  2006\\
\end{flushleft}

\newpage

\section{Introduction}

In a recent series of papers \cite{VW1}--\cite{VW2} we have introduced  a supersymmetric
quantum mechanical matrix model and  studied its rather intriguing properties in the planar approximation, or,
formally,  in the large-$N$, fixed-$\lambda$ limit (here $N$ is the size of our bosonic and fermionic matrices
and $\lambda \equiv g^2 N$ is the usual 't Hooft coupling)\cite{tH}.
In particular, the model exhibits, at a critical value of $\lambda$, $ \lambda_{\rm c} = 1$,
a discontinuous phase transition
 characterized by  the emergence of new supersymmetric vacua on the strong-coupling side of the phase transition, and a consequent jump of Witten's index \cite{WI} across $\lambda_{\rm c}$.

This property was first observed  \cite{VW1, Adriano} in the lowest fermion-number sector of the model, $F=0$ ($F$ being an exactly conserved quantum number for all values of $\lambda$) where {\it one} new supersymmetric vacuum emerges at $\lambda >1$. It was later realized that a similar phenomenon  also occurs at $F=2$ \cite{VW2}, with  {\it two} new supersymmetric vacua popping up at $\lambda >1$. In that same paper
a deeper understanding of this unexpected feature was  gained by considering
the $\lambda \rightarrow \infty$ limit of the model. In this limit the Hamiltonian becomes block-diagonal
in both $F$ and boson number $B$, with blocks of finite size ${\cal N}(F,B)$.

 Furthermore, by combining the strong-coupling limit with some combinatorics arguments \cite{OVW1}, it was conjectured that the pattern found at  $F=2$  should generalize to all even values of $F$:  {\it two} new supersymmetric vacua would occur in each one of these sectors at large $\lambda$.
By going to infinite $\lambda$, and by computing suitable supertraces \cite{OVW1}, we can also  guess which  blocks  should ``hide"  (for a given even $F$) the {\it two}  new ground states:  those with $B = F \pm 1$, a conjecture confirmed by many numerical checks.
Finally, once the infinite-$\lambda$ ground states are identified,  their expression at finite $\lambda$  can be reconstructed  through an explicit formal operation \cite{VW2}, which is expected to lead to a normalizable state if and only if $\lambda > \lambda_{\rm c} $.

In this paper we consider again the $\lambda \rightarrow \infty$ limit, albeit for a different purpose.
We will show that some of the finite blocks (including those where the new supersymmetric vacua occur) can be mapped into (seemingly non-supersymmetric) one-dimensional statistical mechanics models with a finite number of sites and a periodic (cyclic) structure.
More explicitly, we will map some sectors of our model into the XXZ Heisenberg chain with asymmetry parameter
$\Delta = \pm 1/2$. Moreover, it will be argued, and demonstrated numerically, that our system is also equivalent to
   a lattice gas of $q$-bosons in the limit where the quantum-deformation parameter $q$ goes to infinity.

Such kinds of equivalences are not new, the Thirring--sine-Gordon connection \cite{Thirring} being a famous example; however, the equivalence discussed in this paper has two novel features:
\begin{itemize}
\item   It connects a supersymmetric system with a (seemingly)
non-supersymmetric one, hopefully  revealing  hidden supersymmetric features of the latter model.
As an example, the ground state of the XXZ model  with $\Delta = - 1/2$ is known to have amazing (partly proved, partly  conjectured) properties \cite{RS} that could  possibly be explained after realizing that such a ground state is just a supersymmetric vacuum.

\item  It relates a rather abstract quantum mechanical matrix model, in the planar approximation, to some well known statistical system in one-space one-time,  therefore providing
an a priori unexpected physical (string-like?) interpretation of the former.
\end{itemize}

In the next section we recall the definition of our model, its main physical properties, and its
large-$\lambda$ limit. The
equivalence with the XXZ chain and with the $q$-bosonic gas is discussed in the following two sections. We will end with a  summary and a discussion
 of the possible consequences of this equivalence  for the  latter two systems.

\section{A planar supersymmetric matrix model}

The model is simply the $N \rightarrow \infty$ (planar) limit of
an $N \times N$ matrix system
defined by the following supersymmetric charges and Hamiltonian:
\eq
Q= {\rm Tr} [f \ad(1+g\ad)],
\;\;\; \Qd=  {\rm Tr} [\fd (1+g a) a],
\eqx
\eq
H=\{Q^{\dagger},Q\} = H_B+H_F  \, , \label{h1}
\eqx
\eq
H_B= {\rm Tr} [\ad a + g(\ad^2 a + \ad a^2) + g^2 \ad^2 a^2] \, , \label{h2}
\eqx
\eqn
H_F&=& {\rm Tr} [\fd f + g ( \fd f (\ad+a) + \fd (\ad+a) f) \nonumber \\
& + & g^2 ( \fd a f \ad + \fd a \ad f + \fd f \ad a + \fd \ad f a)] \, ,\label{h3}
\eqnx
where bosonic and fermionic destruction and creation operators satisfy
\eq
[a_{ij},\ad_{kl}]=\delta_{il}\delta_{jk}  \,~ ; \,~ \{f_{ij} \fd_{kl}\}=\delta_{il}\delta_{jk} \, ;
\,\,\, i,j,k,l =1,\dots N \, ,
 \label{com}
\eqx
all other (anti)commutators being zero.

Models of this type result from the dimensional reduction of  $D=(1+1)$-dimensional gauge theories.
For example,  two-dimensional Yang--Mills gluodynamics, when reduced to QM,  is described
by a free Hamiltonian
$H_{\rm{YM}_2}=-{\rm Tr}[(a-\ad)^2]/2$ acting on the gauge-invariant subspace of the whole Hilbert
space \cite{CH}. The Hamiltonian (\ref{h1}--\ref{h3}) was designed
to illustrate a new, general method \cite{VW1} of finding the spectrum of gauge systems at infinite number of colours
$N$.
It turned out, however,  to have an interest of its own, by exhibiting  the following non-trivial
properties:
\bi
\item Since (\ref{h3}) conserves the fermionic number $F={\rm Tr}[\fd f]$, the system can be studied separately
for each $F$.
\item The planar model is exactly soluble in the $F=0,1$ sectors, i.e.  the complete energy spectrum and
the eigenstates are  available in analytic form.
\item There is a discontinuous phase transition in the 't Hooft coupling
at $\lambda= \lambda_{\rm c} = 1$. At this point the otherwise discrete spectrum loses its energy  gap and becomes continuous.
\item An  exact duality between the strong and weak coupling phases occurs in the $F=0,1$ sectors.
\item The system exhibits unbroken supersymmetry, i.e. there are exact, SUSY-induced  degeneracies
between bosonic (even $F$) and fermionic (odd $F$) eigenenergies.
\item  In the weak coupling phase, $\lambda < 1$, there exists only one (unpaired) SUSY vacuum. It lies in the
$F=0$ sector and it is nothing else than the empty Fock state $|0\rangle$. For  $\lambda > 1$, however,
the structure is much less trivial and more interesting:  there are {\em two} SUSY vacua in each bosonic sector of the model. For $F=0$ the empty Fock state continues to be a null eigenstate,
but it is joined by another, non-trivial, analytically known ground state. For higher  even $F$, two new
non-trivial vacua appear. This is possible thanks to the rather  intriguing
rearrangement of the members of supermultiplets that occurs across the phase-transition point.
\ei

We have established all these points for the $F=0,1,2,3$ sectors and believe that this structure
persists for arbitrary $F$. This expectation is borne out by considering the infinite
$\lambda \rightarrow \infty$ limit of the Hamiltonian (\ref{h3}). Since this is also the limit in which our Hamiltonian
reveals the above-mentioned  connections to statistical mechanics, let us  recall the strong coupling limit
of our system \cite{VW2} in a little more detail.

Define the (appropriately rescaled) strong coupling SUSY charges by:
\eq
Q_{{\rm SC}}    =  \lim_{\lambda\rightarrow\infty} \frac{1}{\sqrt{\lambda}}~ Q = \frac{1}{\sqrt N} {\rm Tr}(f \ad^2) ~~,~~ Q^{\dagger}_{{\rm SC}}    =  \frac{1}{\sqrt N} {\rm Tr}(\fd a^2)\, .
\eqx
The corresponding strong-coupling Hamiltonian is just their anticommutator.
Doing the algebra carefully, and throwing away terms that do not contribute in the large-$N$ limit, we
find:
\eqn
 H_{{\rm SC}}=\lim_{\lambda\rightarrow\infty} \frac{1}{\lambda} H =
 \frac{1}{N} {\rm Tr}[ \ad^2 a^2 + \fd a f \ad + \fd a \ad f + \fd f \ad a + \fd \ad f a] \, .
\eqnx
 This can be   further simplified with the aid of the  ``planar calculus"  rules  derived in \cite{VW1,VW2};
namely, the third and the fourth terms must be brought to normal form, giving:
\eqn
\fd_{ij} a_{jk} \ad_{kl} f_{li} &=& \fd_{ij}(\ad_{kl}a_{jk} + \delta_{jl}\delta_{kk}) f_{li} \rightarrow N {\rm Tr}[\fd f],\\
\fd_{ij} f_{jk} \ad_{kl} a_{li} &=& \fd_{ij} \ad_{kl} f_{jk}  a_{li} \rightarrow 0\, .
\eqnx
In the above relations we have neglected terms in which the normal ordering does not match the trace-ordering, since such structures do not contribute
to leading order  in $1/N$  \cite{VW1,VW2}.
In conclusion, the strong-coupling Hamiltonian reads:\footnote{To be precise $H_{{\rm SC}}$ contains also a ${\rm Tr}(\fd) {\rm Tr}( f) $ term. It is subleading {\it except} when it acts on\\
$|B=0, F=1 \rangle$  and plays an important role for assuring degeneracy with  $|B=2, F=0 \rangle$.}
\eqn
H_{{\rm SC}}= {\rm Tr}[\fd f +
 \frac{1}{N}(\ad^2 a^2 + \ad \fd a f + \fd\ad f a)].  \label{hstr}
\eqnx
Remarkably, (\ref{hstr}) conserves {\em both} $F$ {\em and} $B={\rm Tr}[\ad a]$. As such,  the infinite Hamiltonian matrix
splits into  {\em finite} blocks labelled by the number of fermionic and bosonic quanta, $(F,B)$.
In each such sector the leading-order (i.e. planar) basis is generated by the single trace of a product of elementary
creation operators. We may thus represent the generic state in a block of given $B$ and $F$ in the form:
\eqn
|m_i,n_i\ra = \frac{1}{{\cal N}_n}
 {\rm Tr}[\ad^{m_1}(\fd)^{n_1}\ad^{m_2}(\fd)^{n_2} \dots (\fd)^{n_k}]|0\ra \, ; \, m_i, n_i  > 0 \, .
\label{plstates}
\eqnx
This is a state with $B=\sum m_i$ bosons and $F = \sum n_i$ fermions, ${\cal N}_n$
being a normalization factor. Owing to the Pauli principle not all configurations of
$\{ m_i,n_i\}$ define a legitimate state.  The detailed rules for counting such states (Pauli-allowed necklaces) follow from  Polya's  theory and
have been developed in \cite{OVW1}: they give the dimensionality of each block of the strong-coupling Hamiltonian (\ref{hstr}). Table 1 shows a map of the first few $(F,B)$ sectors together with their sizes. Since, in the strong-coupling limit, the supersymmetric charges connect states with
the same value of $B+2F$, taking supertraces at fixed $B+2F$ is a way to check whether all states
with such a value of $B+2F$ are paired into supermultiplets or not. Such an analysis was carried out in
\cite{OVW1} and revealed a  bosonic excess by one unit for (and only for) $B+2F = \pm 1(\rm{mod}~ 6)$.

Using the planar rules
 developed in \cite{VW1,Adriano,VW2} the finite-dimensional
Hamiltonian matrix can be readily calculated in each sector. Proceeding in this way, we have discovered
``experimentally" the existence of a ``magic staircase" -- a distinct set of sectors
(labelled by a bold face in Table 1) --  where the zero-energy eigenstates are located
\footnote{The dimensionality of these sectors turns out to be given by Catalan's numbers
 \cite{OVW1}.}. Indeed the magic sectors appear to lie at
\eqn
B=F \pm 1\,\, , \,\,F = 2n \,\,  \Rightarrow B+2F = 3F \pm 1 = 6 n \pm 1~,  \label{mgc}
\eqnx
 in agreement with the supertrace considerations of \cite{OVW1}.  This observation also explains the structure of SUSY vacua
in the whole strong-coupling phase. Indeed,  we were able  \cite{VW2} to give a (formal) way to express SUSY vacua at  finite $\lambda$  in terms of those at $\lambda=\infty$. The connection is expected to lead to normalizable states only at $\lambda >1$.

It remains, however, to understand {\em why} the strong SUSY vacua exist solely in the magic sectors
(\ref{mgc}).  This puzzle finds its solution upon mapping our system into the XXZ Heisenberg chain.

\begin{table}[tbp]
\begin{center}
{
\begin{tabular}{||c|ccccccccccc}
$11\;\;\;$ & 1  &  1 & 6 &  26  & 91 &  273 & 728 & 1768 & 3978 & 8398  & {\bf 16796} \\
$10\;\;\;$ & 1  &  1 & 5 &  22  & 73 &  201 & 497 & 1144 & 2438 & 4862  & 9226 \\
$9\;\;\;$ & 1  &  1 & 5 &  19  & 55 &  143 & 335 & 715 & {\bf 1430} & 2704  & {\bf 4862} \\
$8\;\;\;$ & 1  &  1 & 4 &  15  & 42 &  99 & 212 & 429 & 809 & 1430  & 2424 \\
$7\;\;\;$ & 1  &  1 & 4 &  12  & 30 &  66 & {\bf 132} & 247 & {\bf 429} & 715  & 1144 \\
$6\;\;\;$ & 1  &  1 & 3 &  10  & 22 &  42 & 76 & 132 & 217 & 335  & 497 \\
$5\;\;\;$ & 1  &  1 & 3 &  7  & {\bf 14} &  26 & {\bf 42} & 66 & 99 & 143  & 201 \\
$4\;\;\;$ & 1  &  1 & 2 &  5  & 9 &  14 & 20 & 30 & 43 & 55  & 70 \\
$3\;\;\;$ & 1  &  1 & {\bf 2} &  4  & {\bf 5} &  7 & 10 & 12 & 15 & 19  & 22 \\
$2\;\;\;$ & 1  &  1 & 1 &  2  & 3 &  3 & 3 & 4 & 5 & 5  & 5 \\
$1\;\;\;$ & {\bf 1}  &  1 & {\bf 1} &  1  & 1 &  1 & 1 & 1 & 1 & 1  & 1 \\
$0\;\;\;$ & 1  &  1 & 0 &  1  & 0 &  1 & 0 & 1 & 0 & 1  & 0 \\
\hline 
$B\;\;\;$        &  $  $ & $  $ & $  $
  & $  $ &  $  $ & $  $ & $  $  & $  $ &  $  $ & $  $ & $  $ \\

  $\;\;\;F$        &  $ 0 $ & $ 1 $ & $ 2 $
  & $ 3 $ &  $ 4 $ & $ 5 $ & $ 6 $  & $ 7 $ &  $ 8 $ & $ 9 $ & $ 10 $ \\

   \hline\hline
\end{tabular}
}
\end{center}
\caption{
Sizes of gauge-invariant bases in the $(F,B)$ sectors}
\label{bastab}
\end{table}
\vspace*{1cm}

\section{Equivalence with the XXZ Heisenberg chain}

\subsection{The XXZ model}

The XXZ system \cite{Be,Ba1,Fad} is a one-dimensional
 periodic lattice of size $L$ with a spin $1/2$ variable residing on its sites.
  Its  Hamiltonian depends on an ``asymmetry" parameter $\Delta$ and can be
  written in terms of Pauli's matrices
   (with $\sigma^{\pm} = \frac 12 (\sigma^x \pm i \sigma^y)$)  as:
\eqn
H_{\rm XXZ}^{(\Delta)} &=& -\frac12 \sum_{i=1}^L \left( \sigma_i^x \sigma_{i+1}^x + \sigma_i^y \sigma_{i+1}^y + \Delta~    \sigma_i^z \sigma_{i+1}^z    \right) \nonumber \\
& =& -\sum_{i=1}^L \left( \sigma_i^+ \sigma_{i+1}^- +
 \sigma_i^- \sigma_{i+1}^+  + \frac{\Delta}{2}    \sigma_i^z \sigma_{i+1}^z    \right)
 \equiv -\left(O_{\pm} + \frac{\Delta}{2}   O_z\right) \, ,
\eqnx
where the site $L+1$ is identified with the site $1$. A convenient basis of states,
labelled by two sets of integers $(m_i,n_i )$,  is the following:
\eq
\left| m_i,n_i \right\ra \equiv \left| (0)^{m_1}(1)^{n_1}\dots (0)^{m_r}(1)^{n_r} \right\ra ~,
~ r \ge 1, m_i, n_i > 0
~,~ \sum (m_i +  n_i) = L \, ,
\label{set}
\eqx
where a $(0)^m$ (resp. a $(1)^n$) indicates a sequence of $m$ (resp. $n$) spins pointing down (resp. up). The state is defined up to  cyclic permutations and the permutation giving the smallest binary number can be taken as representative. To the set (\ref{set}) one still  has to add two states: $(0)^L$ and $(1)^L$.

The action of $O_z$ on such states is very simple:
\eq
O_z \left|m_i,n_i\right\ra   = (n_{\rm eq} - n_{\rm opp})~\left|m_i,n_i\right\ra  ~,
\eqx
where $n_{\rm eq}$ ($n_{\rm opp}$) is the number of equal(opposite)-spin  nearest neighbours. It is easy to see that $n_{\rm eq} = L -2r$  while $n_{\rm opp} = 2r$. Therefore:
\eq
O_z \left|m_i,n_i\right\ra   = (L-4r)~\left|m_i,n_i\right\ra  ~.
\eqx
The action of $O_{\pm}$ is only a bit more complicated:
\eq
O_{\pm} \left|m_i,n_i\right\ra   = \sum \left|m'_j,n'_j\right\ra  ~,
\eqx
where the sum extends over the sets $(m'_j,n'_j)$ that are obtained
 from $(m_i,n_i)$ by interchanging, in turn one by one, any pair of opposite-spin  neighbours. As a result, the Hamiltonian commutes with the $z$-component of the total spin and is block-diagonal with blocks of given $m = \sum m_i$ and $ n =\sum n_i~(n+m = L)$.
In conclusion:
\eqn
H_{\rm XXZ}^{(\Delta)} ~~ \left|m_i,n_i\right\ra  &=&
- \frac{\Delta}{2} (L-4r)~\left|m_i,n_i\right\ra -  \sum \left|m'_j,n'_j\right\ra \nonumber \\
 &=&  \left(\frac32 \Delta~ L- 2 \Delta~(L-r) \right) \left|m_i,n_i\right\ra -  \sum \left|m'_j,n'_j\right\ra\, .
\eqnx
It is useful to define a rescaled XXZ Hamiltonian by:
\eqn
\tilde{H}_{\rm XXZ}^{(\Delta)} &\equiv&  -\frac{1}{2\Delta} H_{\rm XXZ}^{(\Delta)} =
 \frac{1}{4} O_z  +   \frac{1}{2\Delta} O_{\pm}
 \nonumber \\
& =& \frac 14 \sum_{i=1}^L \left(   \sigma_i^z \sigma_{i+1}^z +
\frac{2}{ \Delta} ( \sigma_i^+ \sigma_{i+1}^- +
 \sigma_i^- \sigma_{i+1}^+)   \right)\, ,
\eqnx
so that:
\eq
\tilde{H}_{\rm XXZ}^{(\Delta)} ~~ \left|m_i,n_i\right\ra
 =  \left(- \frac34 ~ L + (L-r) \right) \left|m_i,n_i\right\ra +
 \frac{1}{2\Delta} \sum \left|m'_j,n'_j\right\ra \, .
\eqx
For the special  values of the asymmetry parameter $\Delta = \pm \frac12$ we get:
\eq
\tilde{H}_{\rm XXZ}^{(\pm1/2)}   =  - \frac{3}{4} L + H^{(\pm)} \, ,
\eqx
where\footnote{We are grateful to Don Zagier for having pointed out to us this simple version of the XXZ Hamiltonian and for having suggested a possible relation to our matrix model.}:
\eq
H^{(\pm)}~~ \left|m_i,n_i\right\ra  =  (L-r)  \left|m_i,n_i\right\ra \pm \sum \left|m'_j,n'_j\right\ra \, .
\eqx

We will now argue that the strong-coupling Hamiltonian of our supersymmetric matrix model
can be mapped into $H^{(\pm)}$ for some specific values of $B,F$ that include those of the magic staircase.

\subsection{Proof of the equivalence}

Let us start by splitting the strong-coupling Hamiltonian (\ref{hstr}) as follows:
\eq
H_{{\rm SC}}   =  H_{{\rm SC}}^{(d)} +  H_{{\rm SC}}^{(o)}  \label{SC} \, ,
\eqx
where:
\eq
H_{{\rm SC}}^{(d)}   =  {\rm Tr}(\fd f)  +  \frac{1}{ N} {\rm Tr} (\ad^2 a^2 ) \, ,
\label{SCd}
\eqx
\eq
H_{{\rm SC}}^{(o)} =   \frac{1}{ N} \left[ \fd \ad f a + \ad \fd a f  \right] \, ,
\label{SCo}
\eqx
and the labels $d$ ($o$) stand for diagonal (off-diagonal) pieces of the Hamiltonian.

Let us  now  consider the action of $H_{{\rm SC}} $  on the generic
state (\ref{plstates}) whose similarity with the XXZ states (\ref{set}) is evident.
The planar rules for doing that were discussed in \cite{VW1}, \cite{Adriano}.
 For $H_{{\rm SC}}^{(d)}$ we simply find ($F=n$):
\eq
 H_{{\rm SC}}^{(d)} |  n_i,m_i \rangle  = \left(F + \sum_{i=1}^r
(m_i-1) \right) |  n_i,m_i \rangle = (L-r)~  |  n_i,m_i \rangle \, .
 \eqx
The
action of $H_{{\rm SC}}^{(o)}$, on the other hand, is precisely
to interchange each fermion--boson  pair of neighbours, i.e. an
action very close to that of $O_{\pm}$ in the XXZ model.
However, since sign problems can  arise,  we have to treat
various cases separately:

\begin{itemize}

\item  $F$ odd

In this case there are no relative signs from different cyclic orderings in (\ref{plstates}) and therefore the action of
$H_{{\rm SC}}^{(o)}$ on them is exactly the same as that of $O_{\pm}$ on the states (\ref{set}). The relative sign of
$H_{{\rm SC}}^{(d)}$ and $H_{{\rm SC}}^{(o)}$ in $H_{{\rm SC}}$ concides with the one in $H^{(+)}$.
We thus obtain:
\eq
H_{{\rm SC}}   \Leftrightarrow H^{(+)} = \tilde{H}_{\rm XXZ}^{(+1/2)} +  \frac{3}{4} L
 =  - H_{\rm XXZ}^{(+1/2)}   +  \frac{3}{4} L \, .
\eqx
\item  $F$ even and $B$ odd

In this case different cyclic orderings in  (\ref{plstates}) do  carry relative signs.
We will argue that,  by a suitable choice of the basis, we can bring $H_{{\rm SC}}$ to agree, up to a shift, with $ H_{\rm XXZ}^{(-1/2)}$:
\eq
H_{{\rm SC}}   \Leftrightarrow H^{(-)} = H_{\rm XXZ}^{(- 1/2)}   +  \frac{3}{4} L \, .
\eqx
The argument goes as follows:
for the XXZ model let us choose as basis of states:
\eq
(1111\dots1100\dots000), (1111\dots1010\dots000), \dots  \
\label{xxzbasis}
\eqx
The corresponding basis for $H_{{\rm SC}} $ is then taken to be:
\eq
|\fd\fd\fd\fd\dots\fd\fd\ad\ad\dots\ad\ad\ad), - |\fd\fd\fd\fd\dots\fd\ad\fd\ad\dots\ad\ad\ad\rangle, \dots \, ,
\label{scbasis}
\eqx
i.e. we put in correspondence the ones (zeroes) in (\ref{xxzbasis}) with the fermions (bosons) in
(\ref{scbasis}), while assigning a sign $(-1)^k$ to a state in the latter set if it is obtained from the first state  by interchanging $k$ boson--fermion pairs.

It is quite obvious that, on this convenient basis, $H_{{\rm SC}}$ gives the same matrix elements as $H^{(-)}$, modulo the possibility that $H_{{\rm SC}} $ produces a cyclic permutation
of a state in the above basis. However, even in this case, the correspondence works fine thanks to the fact that the relative sign $(-1)^p$ originating from Fermi statistics (where $p$ is the number of fermions to be interchanged) is equal to the relative sign $(-1)^k$ counting the number of boson--fermion interchanges. This is so since  taking a fermion from the last to the first entry corresponds to an odd number of fermion--fermion {\it and} boson--fermion interchanges, while doing the same with a boson involves an even number of interchanges of each type.
Note that for this to be true it is essential that $B$ be odd and $F$ be even, and, therefore, that $L$ be odd.
The importance of $L$ being odd was much emphasized in \cite{RS}.

\item $F$ and $B$ even

In this case we have found no simple relation between $H_{{\rm SC}}$ and $H_{\rm XXZ}$ for any choice
 of $\Delta$.

\item We may add here a side remark: in the case of $F$ odd, if we define a new (non-supersymmetric) Hamiltonian:
\eq
\tilde{H}_{{\rm SC}}   =  H_{{\rm SC}}^{(d)} -  H_{{\rm SC}}^{(o)} \, ,
\eqx
we also find:
\eq
\tilde{H}_{{\rm SC}}   \Leftrightarrow  H^{(-)} = H_{\rm XXZ}^{(- 1/2)}   +  \frac{3}{4} L \, .
\eqx
\end{itemize}

\subsection{Implications of supersymmetry on the XXZ model}

The final form of our equivalences reads:
\eqn
H_{{\rm SC}}(F,B)=\left\{\begin{array}{ccc}
             -H_{\rm XXZ}^{(+ 1/2)}+\frac{3}{4}L \ ,\   & F ~{\rm  odd}, &\\
             +H_{\rm XXZ}^{(- 1/2)}+\frac{3}{4}L \ , \  & F ~{\rm  even}, B~ {\rm odd} \, .
            \end{array}
            \right.              \label{xxzeq}
\eqnx
 Eq.(\ref{set}) implies that the parameters of both systems are  related as follows:
\eqn
L & = & F + B , \\
S_z & = & \frac 12(F - B) \, ,  \label{fbvs}
\eqnx
where $S_z$ is the conserved component of the total spin.  In addition,
the spectrum on the spin side should be computed in the sector that is invariant under the lattice shifts.

We checked Eqs.(\ref{xxzeq}) for all sectors with $ 5 \le F+B \le 9 $.
 Everything works as expected, including the magic sectors
with a zero eigenvalue. Since the spectrum of $H_{{\rm SC}}$ is positive semi-definite,
the 2nd of Eqs. (\ref{xxzeq}) gives a simple and elegant proof
that, for $L$ odd,
the states considered in \cite{RS} are indeed ground states of the XXZ chain.
The fact that  they have $S_z = \pm \frac 12$
just corresponds to our SUSY ground states having $B= F \pm1$ and $F$ even.
Moreover, for sectors with $F$  {\em and} $B$ even, where the equivalence
is not expected to work, we indeed find different spectra in the two models.

Supersymmetry also implies that the (non-vanishing) spectrum of $H_{{\rm SC}}$ in the sector
$(F,B)$ has to be contained
in the spectra of the ``neighbouring" sectors with $(F\mp1,B\pm2)$. If we take
$B$ odd, $B \pm 2$ is also odd and therefore we are always in a situation in
which we are able to connect $H_{{\rm SC}}(F,B)$ to $H_{\rm XXZ}(n,m)$. Obviously,
$n=F$ and $m=B$ are conserved by  $H_{\rm XXZ}$ as well as by $H_{\rm SC}$.
After some trivial algebra, we get the following predictions
(with integer  $\mu$ and $\nu$ ensuring even/odd $m$ and $n$ respectively):

\vspace{2mm}

{\bf 1. The  spectrum of $H_{\rm XXZ}^{(+1/2)}(2\nu +1, 2\mu + 1)$ is contained
in the combined spectra of $-H_{\rm XXZ}^{(-1/2)}(2\nu,2\mu +3)-3/4 $
and $-H_{\rm XXZ}^{(-1/2)}(2\nu+2,2\mu - 1)+ 3/4 $},

and:

{\bf 2. The excited spectrum of $H_{\rm XXZ}^{(- 1/2)}(2\nu, 2\mu+ 1)$
is contained in the combined spectra of $-H_{\rm XXZ}^{(+1/2)}(2\nu - 1,2\mu + 3)+3/4$
and $-H_{\rm XXZ}^{(+1/2)}(2\nu +1,2\mu -1)-3/4 $.}
\vspace{2mm}

Since the $H_{\rm XXZ}(n,m)$ Hamiltonian is symmetric under
the exchange of $m$ and $n$, all variations of the above relations resulting from
the interchange  $m \leftrightarrow n$ are also valid.
We made extensive numerical checks of these (to our knowledge novel) relations between different
XXZ models at different values of $\Delta$.

\subsection{Bethe ansatz solutions for the lower sectors \\ of the XXZ chain}

The XXZ chain is integrable \cite{Fad}. In particular
the eigenenergies of $H_{\rm XXZ}(\Delta)$ are given exactly by the Bethe ansatz \cite{Be,XXZ}:
\eqn
E_{\rm XXZ}(\Delta)=-L\frac{\Delta}{2} + 2 m \Delta -2 \sum_{j=1}^m \cos{p_j}, \label{baen}
\eqnx
where the momenta $ -\pi < p_j < \pi $ satisfy the following set of Bethe equations:
\eqn
e^{i L p_j} = (-1)^{m-1}  \prod_{l=1,l\neq j}^m  e^{i(p_j-p_l)}
\frac{e^{i p_l}+e^{-i p_j} - 2\Delta}{e^{i p_j}+e^{-i p_l} - 2\Delta},\;\;\; j=1,\dots,m. \label{beqs}
\eqnx
With $m$ denoting the number of down spins in a chain. For the supersymmetric model this translates into
\eqn
\label{SBAodd}
E_{{\rm SC}}=&F+2\sum_{j=1}^B \cos{p_j},& \;\;\;{\rm for}\; F\;\;{\rm odd}~,~ \Delta = + \frac12 \\
E_{{\rm SC}}=&F-2\sum_{j=1}^B \cos{p_j},& \;\;\;{\rm for}\; F\;\;{\rm even},\;\;{\rm and}\;\;\ B\;\; {\rm odd}
~,~ \Delta = -  \frac12 \, .
\label{SBAeven}
\eqnx
Given the Bethe momenta $p_i$, all eigenvectors can also be  explicitly constructed.
The whole problem reduces therefore to the solution of the non-linear equations (\ref{beqs}).
Consequently, the existence of the magic staircase with its supersymmetric vacua follows directly
from the Bethe ansatz solution of the XXZ chain. The literature on the latter subject
is huge, see e.g. \cite{Fad}--\cite{Ku} for more references.
We shall content ourselves here  with  formulating only an  ``ansatz within the Bethe ansatz",
which reduces the number of Bethe momenta needed to find our SUSY vacua. Solving numerically
Eqs. (\ref{beqs} ) for a few low-$F$ magic sectors, we have found that the zero-energy momenta satisfy (see Table 2):
\eqn
p_1=0,\;\;\; p_{2k+1}=-p_{2k}, \;\;\; k=1,...,(B-1)/2.    \label{an2}
\eqnx
In words, there is always one zero momentum, and the remaining ones come in pairs with
opposite sign\footnote{Recall that the magic sectors only occur  for odd values of  $B$.}. We conjecture that this configuration gives the zero-energy state for arbitrary $B=F\pm 1$ and even $F$.

\begin{table}[h]
\begin{center}
{
\begin{tabular}{cc|cccc}
\hline\hline
$(F,B)$ & ${\cal N}(F,B)$ &$p_1/\pi$  &  $p_2/\pi$ & $p_4/\pi$  &  $p_6/\pi$    \\
\hline\hline
$(4,3)$ & 5 &$0.0$  &  $1/3$ &   &     \\
$(4,5)$ & 14&$0.0$  &  $0.260669$ & $0.558585$  &      \\
$(6,5)$ & 42&$0.0$  &  $0.210767$ & $0.432222$  &      \\
$(6,7)$ &132&$0.0$  &  $0.178899$ & $0.364275$  &  $0.583645$    \\
\end{tabular}
}
\end{center}
\caption{
Bethe momenta of the supersymmetric vacua in some of the lowest
magic $(F,B)$ sectors, see Table 1}
\label{bastab}
\end{table}
\vspace*{1cm}

It is perhaps amusing that the ansatz (\ref{an2}) allows the Bethe phase
factors to  be derived for the first three vacua listed in Table 2 in analytic form. Assuming that they correspond to
zero-energy
eigenstates, we look for simultaneous solutions of Eqs. (\ref{beqs}) and $E(F,B)=0$. This problem
can be solved algebraically. Defining $x_i = e^{i p_i}$, we find:
\eqn
\lefteqn{\bm{ F=4,\;\;B=3:}}  && \nn \\
x_2&=&\frac{1}{2}(1+ i \sqrt{3})\nn \\
\lefteqn{\bm{F=4,\;\;B=5:}} && \nn \\
x_2&=&\frac{1}{64}\left(16-i \sqrt{2}\sqrt{15+\sqrt{33}}(7-\sqrt{33})
+4\sqrt{-16(3+\sqrt{33})+i 2\sqrt{2}\sqrt{15+\sqrt{33}}(9+\sqrt{33})}\right)\nn  \\
x_4&=&\frac{1}{64}\left(16+i \sqrt{2}\sqrt{15+\sqrt{33}}(7-\sqrt{33})
-4\sqrt{-16(3+\sqrt{33})-i 2\sqrt{2}\sqrt{15+\sqrt{33}}(9+\sqrt{33})}\right) \nn \\
\lefteqn{\bm{F=6,\;\;B=5:}} && \nn \\
x_2&=&\frac{1}{72}\left(36+i \sqrt{2}\sqrt{11+\sqrt{13}}(7+\sqrt{13})
+6\sqrt{2}\sqrt{6(-3+\sqrt{13})+i \sqrt{2}\sqrt{11+\sqrt{13}}(-5+\sqrt{13})}\right)\nn \\
x_4&=&\frac{1}{72}\left(36+i
\sqrt{2}\sqrt{11+\sqrt{13}}(7+\sqrt{13})
-6\sqrt{2}\sqrt{6(-3+\sqrt{13})+i
\sqrt{2}\sqrt{11+\sqrt{13}}(-5+\sqrt{13})}\right) \nn \label{xy45} \, ,
\eqnx
 corresponding to  the algebraic representations of the phase factors
given in Table 2. Although it is not self-evident, it can be proved
algebraically that the above numbers have modulus 1.

\subsection{The Razumov--Stroganov states and supersymmetry}

Interestingly, the magic staircase, with its supersymmetric vacua, connects directly to properties of
one-dimensional spin chains. Beginning with
the classic paper of Baxter \cite{Ba1}, some simple eigenvalues of
the XXZ Heisenberg chain were discovered. In particular, Baxter has shown the existence
of a ground state with $S_z = \pm \frac{1}{2}$ and  energy
$-\frac34 L$ for infinite $L$ and $\Delta=-\frac{1}{2}$.
More recently, Razumov and Stroganov \cite{RS}  extended this result to any finite, odd $L$ and made several conjectures on the properties of the eigenvector that corresponds to the above-mentioned ground state (for recent developments see, e.g.  \cite{Zuber}).

   It follows directly from (\ref{xxzeq}) and (\ref{fbvs}) that these states are nothing
   but the supersymmetric
vacua of our planar model with $B=F \pm 1, F\; even$. Hence  Razumov and Stroganov's
above-mentioned result  guarantees the
existence of one bosonic SUSY vacuum in each one of our magic sectors.
Hopefully, this hidden supersymmetry will help understanding
and/or proving the other amazing (and so far mostly conjectured) properties of Baxter's ground states.

It is tempting to say that the two families of vacua, i.e. those
with some given even $F$ and $B=F\pm 1$,  are related by the
usual inversion, $\sigma_i \rightarrow -\sigma_i$, symmetry.
However, this transformation has to be {\em coupled} with a change in the lattice
size $L \rightarrow L+2$. This brings the issue of whether  a
change of $L$ should be considered as a symmetry. At first sight this
 looks  a little premature, and in fact, for the above vacuum
solutions, it is pure semantics. However, this is no longer the case  when we
consider the implications of supersymmetry on the whole
spectrum, including all excited states. The supersymmery generators
act as  $F \rightarrow F \pm 1$ and $B\rightarrow B \mp 2$,
corresponding to $L\rightarrow L\mp 1$. Therefore they do relate the
spectra of excited states on {\em different} lattices: the XXZ
chain turns out to have a hidden supersymmetry with different
members of its dynamical supermultiplets living on different lattices!

\section{Equivalence with a $q$-bosonic gas}
Surprisingly, our infinite-coupling planar system is exactly equivalent to yet another,
and apparently
entirely different,  model.
To expose this equivalence we use the original labelling of planar states with $F$ fermions
\eqn
|n_1,n_2,\dots,n_F\ra = \frac{1}{{\cal N}_n} {\rm Tr}[\ad^{n_1}\fd\ad^{n_2}\fd\dots \ad^{n_F}\fd]|0 \ra \  .
\label{plstate2}
\eqnx
Consider now the action of the strong-coupling Hamiltonian,
\eqn
H_{{\rm SC}}= {\rm Tr}[\fd f +
 \frac{1}{N}(\ad^2 a^2 + \ad \fd a f + \fd\ad f a)] \, ,  \label{hstr2}
\eqnx
on a state (\ref{plstate2}). Using the planar rules developed in \cite{VW1}--\cite{VW2},
it is easy to show that the first two terms do not change the initial state and give rise to the
following diagonal elements:
\eqn
\la n_1,n_2,\dots,n_F | H_{{\rm SC}} |n_1,n_2,\dots,n_F\ra = F + \sum_{i=1}^F (n_i -1 +  \delta_{n_i,0})  =  B + \sum_{i=1}^F \delta_{n_i,0} \, .
\eqnx
Using the same planar rules, the remaining matrix elements can be easily obtained  from the action of the last two terms, e.g.
\eqn
  {\rm Tr}[\ad \fd a f]|n_1,n_2,\dots,n_F\ra =\frac{{\cal N}_{n_1}}{{\cal N}_n}|n_1-1,n_2,\dots,n_F+1\ra+
  \frac{{\cal N}_{n_2}}{{\cal N}_n}|n_1+1,n_2-1,\dots,n_F\ra+ \nn \\
\frac{{\cal N}_{n_3}}{{\cal N}_n}|n_1,n_2+1,n_3-1,\dots,n_F\ra+
\dots+\frac{{\cal N}_{n_F}}{{\cal N}_n}|n_1,n_2,\dots,n_{F-1}+1,n_F-1\ra \, , \nn \\
\label{Hsof1}
\eqnx
i.e. this operator annihilates one bosonic quantum in a group $i$ and adds one at $i-1$,
meaning at the cyclic left of $i, i=1,...,F$. Similarly the last term moves one quantum
to the cyclic right of $i$.
Here, ${\cal N}_{n} ~ ({\cal N}_{n_f})$ are the normalization factors of the initial (final) states of our basis.
They contain some powers of $N$ and degeneracy factors $d_s$. When calculating matrix elements
of the Hamiltonian $H_{{\rm SC}}$, all $N$ factors cancel with the $1/N$ in (\ref{hstr2}) and we are left
only with the square roots of the ratios of corresponding $d_s$ factors.

It turns out that this
Hamiltonian  also describes the following system.
Consider a one-dimensional, periodic lattice of length $F$.
Put at each lattice site a bosonic degree of freedom described by the creation/annihilation (c/a)
 operators $a_i$, $i=1,...,F$ and use
 the harmonic oscillator basis. The states
$|n_1,n_2,...,n_F\ra$
are described by the configuration of $F$ integer occupation numbers as before.

The new Hamiltonian reads
\eqn
H=B + \sum_{i=1}^F \delta_{N_i,0} + \sum_{i=1}^F b_i \bd_{i+1} + b_i \bd_{i-1},   \label{hlat}
\eqnx
where $N_i$ is the usual operator of the number of quanta at a site $i$, $N_i=\ad_i a_i $.
In words: the second term counts the number $n_{zer}$ of empty (unoccupied) sites in a given basis state and
returns this state multiplied by $n_{zer}$.
$B$ is the total number of bosonic quanta, $B=n_1+n_2+...+n_F$.
The  $\bd_i$ ($b_i$) operators create (annihilate) one quantum {\em without}
the usual $\sqrt{n}$ factors.
Omitting momentarily the index $i$:
\eqn
\bd|n\ra = |n+1\ra,\;\;\; b|n\ra = |n-1\ra,\;\;\; b|0\ra\equiv 0 \, . \label{bvac}
\eqnx
In terms of the usual $a, \ad$ operators they read:
\eqn
\bd=\ad \frac{1}{\sqrt{N+1}},\;\;\;\;\;b=\frac{1}{\sqrt{N+1}}a,\;\;\;\; {\rm and}\;\;\; b|0\ra \equiv 0 \, ,
\eqnx
where again $N=\ad a$. The $b$ operators have  non-standard commutation relations:
\eqn
[b,\bd]=\delta_{N,0} \, .  \label{nscom}
\eqnx

This Hamiltonian conserves the total bosonic number, as before.
It is also invariant under lattice shifts and, consequently, commutes with the lattice-shift
operator $U$ defined as:
\eqn
U|n_1,n_2,...,n_F\ra=|n_2,n_3,...,n_F,n_1 \, .\ra
\eqnx
Therefore,  the Hilbert space of states with fixed $B$ can be further split into sectors with fixed
eigenvalues of $U$:
\eqn
\lambda_U^{(m)}=e^{i m \frac{2\pi}{F} },\;\;m=1,2,...,F \, .
\eqnx
We claim that the spectrum of the above $H$, in the sector with $\lambda_U = (-1)^{F-1}$, exactly coincides with our spectrum of $H_{{\rm SC}}$, {\em for all} $F$ and $B$.

The main steps in understanding  this equivalence are as follows:
\bi
\item Our planar states (\ref{plstate2}) are defined modulo $Z_F$ shifts. Without
fermionic degrees of freedom this would be taken care of by
requiring $\lambda_U = 1$ for the bosonic system. The minus sign is the consequence of the
fermionic operators in (\ref{plstate2}):  under the $Z_F$ shifts they acquire the phase
$(-1)^{F-1}$.
\item For even $F$, the projection for the $\lambda_U=-1$ sector plays another important role. Namely,
it removes states that are not allowed by the Pauli principle.

\item Finally, the degeneracy factors required in (\ref{Hsof1}) are correctly taken into account
by the linear combinations corresponding to the condition $\lambda_U=(-1)^{F-1}$.
\ei

The last point is particularly non-trivial. We have therefore double-checked this equivalence by
diagonalizing both Hamiltonians in a range of sectors : $3 \le F \le 7,\;\;\; 3 \le B \le 7$.
All spectra are indeed identical,  including again the supersymmetric vacua in the magic sectors.

Notice that this second equivalence also works  for the ``bad" sectors with both $F$ and $B$ even.
These are the only sectors where the Pauli principle is effective and eliminates some of the
planar states. This is why  there is no XXZ equivalence in these sectors.
Nevertheless, because of the $\lambda_U=-1$ projection, the bosonic-model equivalence applies to
these cases as well.

Interestingly,  the system of ``funny" $b$ and $\bd$  operators turns out to be a
particular limit of the so-called
$q$-deformed harmonic oscillator algebra, well  known in the literature \cite{QB1,QB2}.
The transitions (\ref{bvac}) (without the square roots) are referred to as {\em assisted} transitions.

The $q$-boson  operators satisfy (for one degree of freedom)
\eqn
b \bd - q^{-2} \bd b =1,\;\;\;[N,b]=-b,\;\;\;[N,\bd]=\bd \, ,
\eqnx
with $N=\ad a$, the usual occupation number operator. The
$b,\bd$ operators are related to the standard $a,\ad$ by
\eqn
b=\sqrt{\frac{[N+1]_q}{N+1}}a,\;\;\;\bd=\ad \sqrt{\frac{[N+1]_q}{N+1}},
\eqnx
where:
\eqn
[x]_q\equiv \frac{1-q^{-2x}}{1-q^{-2}} \, .
\eqnx
An  alternative form of the commutation relations reads:
\eqn
[b,\bd]=q^{-2N} \, .
\eqnx
It can readily be checked  that the usual harmonic oscillator algebra is recovered in the limit
$q\rightarrow 1$, $b,\bd \rightarrow a,\ad$.
On the other hand, in the limit $q\rightarrow \infty$, $b$ and $\bd$ become our c/a operators satisfying
(\ref{bvac})--(\ref{nscom}).

An important point is that the $q$-bosonic Hamiltonian without the commutator (or $\delta$) term in (\ref{hlat})
is exactly soluble for all values of the deformation parameter $q$.
However,  with the additional $\delta$ term, it is not.
On the other hand, given the present chain of equivalences, we see that the above non-linear
system of $q=\infty$-bosons {\em is} soluble in terms of the Bethe ansatz for the XXZ chain.
Finally, and similarly to the latter  case, the equivalence we observed  exposes a hidden,
unbroken supersymmetry of $\infty$-bosons with supersymmetric partners living on lattices
of different sizes.

\section{Discussion}

This article is the third in a series studying the quantum mechanics of a simple
supersymmetric matrix model.  Designed originally to illustrate the
usefulness of the large-$N$ approximation directly in terms of a Hamiltonian and a  Hilbert space, the model turned out to have a very rich physics by itself, as amply illustrated in  \cite{VW1}--\cite {VW2}.

Here we have uncovered an even more intriguing aspect of this model: its connection, in the
infinite 't Hooft-coupling limit, to one-dimensional statistical  systems
in which supersymmetry, if present,
is very well concealed.
We have found that our supersymmetric planar model  is exactly equivalent
to {\em two} such systems:  the quantum
XXZ Heisenberg chain, and a lattice gas of $q$-bosons.

The intriguing pattern of  strong coupling vacua discovered in the matrix
model finds its explanation in the unusually simple ground states of
the XXZ chain found by Baxter more than thirty years ago.
Vice versa, the XXZ chain turns out to have a hidden
supersymmetry, which explains a host of degeneracies between seemingly unrelated
energy eigenstates. Interestingly, the supersymmetry transformations change the
number of lattice sites, so that different members of the supermultiplets live
on different lattices.
But even within a single sector/lattice, supersymmetry may turn out to be a powerful tool for studying
the properties of the ground state and for understanding the meaning of the RS conjectures \cite{RS}.
The fact that such  ground states are annihilated by two supercharge operators should imply distinct
properties for them, while finding an operator with the right algebra would allow  to
generate the full ``staircase" of ground states starting from the lowest and simplest ones.

The second system, a gas of $q$-bosons in the limit of an infinite deformation parameter $q$,
is equivalent to our matrix model to an even greater degree than the Heisenberg chain.
While the equivalence with the XXZ model holds only for a subset of all
sectors of the Hilbert space, the $\infty$-bosonic gas is equivalent in all sectors
of conserved boson number and for all lattice sizes.  To the best of our knowledge, the specific Hamiltonian of the bosonic gas was unsolved till now; but  in view of
our chain of equivalences, the system should turn out to be actually soluble (e.g.
via the Bethe ansatz for the XXZ chain).

Finally,  exact solubility of the XXZ model directly implies that same property for
our large-$N$, supersymmetric matrix model  at
infinite 't Hooft coupling and, therefore, in ``three quarters" of its fermionic sectors (since we have found no correspondence for the even-$B$, even-$F$ sectors).
There are also indications that this solubility can be extended to the whole
strong-coupling phase using the technique developed in \cite{VW2}. The reverse is also an interesting question: are there statistical systems that can be mapped into our matrix model at {\it finite} 't Hooft coupling?

Recently relations between the field theories and spin chains have become
the subject of much interest
and excitement in connection with the AdS/CFT correspondence, which
opens a new way of studying gauge theories \cite{MZ}-\cite{Ro}\footnote{We recall that, so far,
the  easiest applications of the AdS/CFT correspondence are also made in the  $\lambda \rightarrow \infty$ (i.e. supergravity) limit.}. In particular,
the mapping between the
${\cal N}=4$ supersymmetric Yang--Mills theory and the XXZ chain with
$\Delta=1/2$ has been discussed
by Belitsky et al. \cite{BK}. It is perhaps
not an accident that we find
a similar gauge--spin relation in our much simpler model.

\section*{Acknowledgements}
We wish to thank D. Zagier for calling our attention on the Razumov-
Stroganov conjectures and E. Onofri for several instructive discussions.
GV would like to thank P. Di Vecchia for useful discussions and encouragement.
JW thanks H. Arodz, A. Gorsky, G. Korchemsky and R. Roiban for discussions,
and the Benasque Centre for Science
for its hospitality. This work is partially supported
by the grant of Polish Ministry of Science and Education
P03B 024 27 (2004)--(2007).

\section*{Note added in proofs} 
After this paper had been submitted we were informed by Jan de Gier that the existence of a hidden supersymmetry in the XXZ spin chain had already been pointed out in Ref. \cite{FNS} and further studied in later work (e.g. in \cite{SUSY}). It looks however that, while  in those papers supersymmetry is realized non-linearly just in terms of fermionic variables, in our case it takes a simpler form in terms of an equal number of bosonic and fermionic degrees of freedom. We are grateful to Dr. de Gier for this information.

\end{document}